\documentclass[fleqn,10pt]{wlscirep}
\usepackage{epsfig}
\usepackage{graphicx}
\usepackage{float}
\usepackage{lipsum}
\usepackage{array}
\usepackage[normalem]{ulem}
\usepackage{color}

\title{Circuit-Based Quantum Random Access Memory for Classical Data}

\author[1,2]{Daniel K. Park}
\author[1,2,3,4]{Francesco Petruccione}
\author[1,2,3,*]{June-Koo Kevin Rhee}

\affil[1]{School of Electrical Engineering, KAIST, Daejeon, 34141, Republic of Korea}
\affil[2]{ITRC of Quantum Computing for AI, KAIST, Daejeon, 34141, Republic of Korea}
\affil[3]{Quantum Research Group, School of Chemistry and Physics,
University of KwaZulu-Natal, Durban, 4000, South Africa}
\affil[4]{National Institute for Theoretical Physics, KwaZulu-Natal, Durban, 4000, South Africa}

\affil[*]{rhee.jk@kaist.edu}

\keywords{Quantum machine learning, Quantum database, Quantum random access memory}

\begin{abstract}
A prerequisite for many quantum information processing tasks to truly surpass classical approaches is an efficient procedure to encode classical data in quantum superposition states. In this work, we present a circuit-based flip-flop quantum random access memory to construct a quantum database of classical information in a systematic and flexible way. For registering or updating classical data consisting of $M$ entries, each represented by $n$ bits, the method requires $O(n)$ qubits and $O(Mn)$ steps. With post-selection at an additional cost, our method can also store continuous data as probability amplitudes. As an example, we present a procedure to convert classical training data for a quantum supervised learning algorithm to a quantum state. Further improvements can be achieved by reducing the number of state preparation queries with the introduction of quantum forking.
\end{abstract}
\begin{document}
\flushbottom
\maketitle

\section*{Introduction}

The theory of quantum information processing promises to accelerate certain computational tasks substantially. In practice, the computational cost of generating an arbitrary quantum input state~\cite{Mottonen.QIC.2005,PhysRevA.64.014303} must be addressed to ensure the speedup. The ability to efficiently convert classical data into quantum states is essential in many algorithms with complex data sets, such as quantum searching~\cite{Grover:1996:FQM:237814.237866}, collision finding~\cite{10.1007/BFb0054319}, and quantum Fourier transform~\cite{QCandQI}. The demand for such ability has continued to grow with recent discoveries of quantum algorithms for data analysis and machine learning applications with large classical data~\cite{QMLreview,qPCA,PhysRevLett.113.130503,PhysRevLett.103.150502,qClassificationEPL}. Quantum simulation also requires the preparation of a quantum register in the initial physical state of the simulated system~\cite{PhysRevA.97.052329}. One promising avenue is to use quantum random access memory (QRAM)~\cite{QRAMPhysRevLett.100.160501}, a device that stores either classical or quantum data with the ability to query the data with respect to superposition of addresses. The bucket brigade (BB) model for QRAM proposed in refs.~\cite{QRAMPhysRevLett.100.160501,QRAMPhysRevA.78.052310} requires $O(\log_2 (M))$ address qubits, $O(M)$ qutrits for routing, and $O(M)$ classical or quantum memory cells for $M$ binary data. The content of multiple data cells can be returned in superposition with only $O(\log_2 (M))$ qutrits activated after $O(\log_2^2 (M))$ time steps. A critical assumption for the practicality of this scheme is that the inactive routing elements do not render noticeable errors.

Since quantum operations are applied directly to the qubits that form the QRAM as the state preparation is followed by a quantum algorithm, it is favorable to build a QRAM based on the quantum circuit model. At the same time, a QRAM should be a good interface to classical data for big data applications. In this work, we propose flip-flop (FF) QRAM, which is constructed with the standard circuit-based quantum computation and without relying on a routing algorithm. The FF-QRAM can read unsorted classical data stored in memory cells, and superpose them in the computational basis states with non-uniform probability amplitudes to create a specific input state required in a quantum algorithm. Also, the classical information stored in the quantum state can easily be updated with the same FF-QRAM process.
The cost for writing or modifying $M$ classical data represented as $D=\{(\vec d^{(l)},b_l)|0\leq l <M\}$, where $\vec{d}^{(l)}$ represents $n$ bits of information and $b_l$ is the attribute of $\vec{d}^{(l)}$, is $O(n)$ qubits and $O(Mn)$ quantum operations that are commonly found in many known algorithms. The probability amplitudes can be modified by post-selection at an additional cost of repeating the process and single qubit measurement.
In addition, the FF-QRAM architecture can serve as a building block for the classical-quantum interface. 

A quantum state prepared by QRAM as an input to a specific algorithm cannot be reused once measured (the quantum measurement postulate), nor be copied (the no-cloning theorem) for another task. Thus, in general, the QRAM cost seems unavoidable per algorithm run, even when performing a set of algorithms with an identical input state. Here we introduce a process of quantum forking inspired by process forking in computer operating systems, which creates a child process that can evolve independently~\cite{UnixForking}. Quantum forking is a framework to split unitary processes in superposition, with which the number of QRAM queries can be reduced in certain applications.

\section*{Results}

A quantum superposition state prepared with respect to classical data for quantum computation can be referred to as quantum database (QDB). In the most general form, a QDB can be expressed as
\begin{equation}
\label{eq:psiqdb}
|\psi\rangle_\text{QDB} = \sum_{l=0}^{M-1} b_l |\vec d^{(l)}\rangle,
\end{equation}
where $\vec d^{(l)}=d_0^{(l)}d_1^{(l)}\ldots d^{(l)}_{n-1}\in\lbrace 0,1\rbrace^n$ denotes a string of quantum bits in the computational basis to represent a classical data, and $M$ is the number of data. For big data applications, a typical quantum representation of $\vec d^{(l)}$ can be decomposed as $|\vec d^{(l)}\rangle = |\vec x^{(l)}\rangle|l\rangle$, where $\vec x^{(l)}$ and $l$ denote a data entry and the corresponding label, respectively~\cite{qPCA, qClassificationEPL}. The probability amplitude, $b_l$, can encode continuous data as required in some applications~\cite{qClassificationEPL,PhysRevX.7}, or represent the normalized occurrence of the data entry $\vec x^{(l)}$. For $M$ numbers of $n$-bit data, $n+m$ qubits are sufficient to realize the QDB, where $n$ and $m=\lceil\log_2(M)\rceil$ qubits encode the data and the label, respectively. The number of label qubits can be reduced by labelling only the data that appears more than once. The label qubits are unnecessary when all data entries, $\vec x^{(l)}$, are unique.

\subsection*{Flip-flop QRAM}

The FF-QRAM is used to generate a QDB as follows. Consider a quantum computer with an $(n+m)$-qubit bus state to encode a big data class database. The bus qubit state can be arbitrary, and it defines which computational basis states, $|j\rangle_B$, are accessed with probability amplitudes, $\psi_j$. A QRAM operation on the bus qubit superposes a set of classical data $D$ as
\begin{equation}
\label{eq:qram}
\text{QRAM}(D)\sum_{j} \psi_j |j\rangle_B |0\rangle_R \equiv \sum_{l}\psi_l | \vec d^{(l)}\rangle_B |b_l\rangle_R ,
\end{equation}
where the subscript $B$ ($R$) indicates the bus (register) qubit, and the register qubit can include the probability amplitudes for encoding the analog data. The FF-QRAM is implemented systematically with standard quantum circuit elements, which include the classically-controlled Pauli $X$ gate, $\bar c X$, and the $n$-qubit controlled rotation gate, $C^{n}R_p(\theta)$. The $\bar c X$ flips the target qubit only when the classical control bit is zero. The $C^{n}R_p(\theta)$ gate rotates the target qubit by $\theta$ around the $p$-axis of the Bloch sphere only if all $n$ control qubits are $1$.

\begin{figure}[t]
\centering
\includegraphics[width=0.7\linewidth]{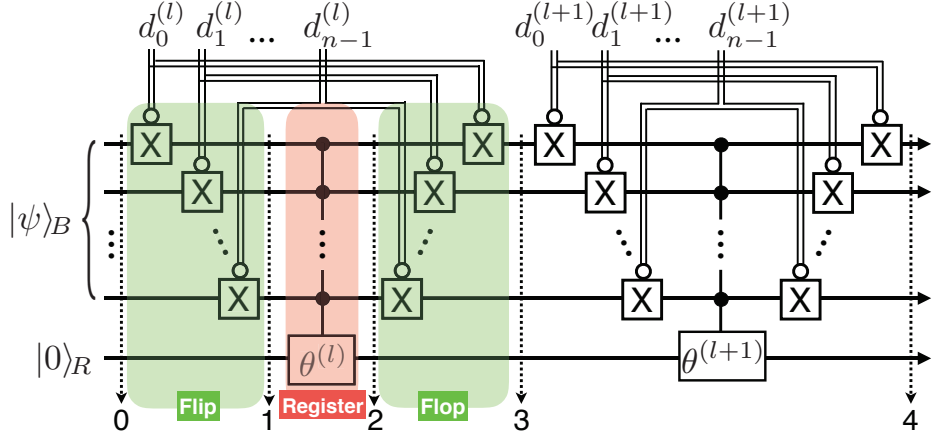}
\caption{Quantum circuit for FF-QRAM that writes bit strings $\vec d^{(l)}$ and $\vec d^{(l+1)}$ as a quantum superposition state with probability amplitudes using multi-qubit controlled rotation gates determined by $\theta^{(l)}$ and $\theta^{(l+1)}$, respectively. The double lines indicate classically controlled operations, and the empty (filled) circle indicates that the gate is activated when the control bit (qubit) is 0 (1). The dotted and numbered arrows indicate the various steps described in the main text.}
\label{fig:qRAM_fig1}
\end{figure}

The underlying idea of the FF-QRAM model is depicted in Fig.~\ref{fig:qRAM_fig1}, describing the procedure to superpose two independent bit strings $\vec d^{(l)}$ and $\vec d^{(l+1)}$ with target probability amplitudes in the bus qubit state, $|\psi\rangle_B$. In this example, the label qubits, $|l\rangle,$ are omitted without loss of generality to deliver the main idea.
The initial state can be expressed with a focus on $\vec d^{(l)}$ as
\begin{equation}
\label{eq:psi0}
|\psi_0\rangle_l=\psi_{\vec d^{(l)}}|\vec d^{(l)}\rangle|0\rangle_R+\!\!\sum_{j\neq \vec d^{(l)}}\!\!\!\! \psi_j |j\rangle|0\rangle_R,  
\end{equation}
where $|\psi_s\rangle_l$ denotes the state of the $(n+1)$ qubits in the process of writing the $l$th data entry, observed at the $s$th step in Fig.~\ref{fig:qRAM_fig1}. The $\bar c X$ gates controlled by $\vec d^{(l)}$ rearrange the computational basis states of the bus qubit so that $|\vec d^{(l)}\rangle$ becomes $|1\rangle^{\otimes n}$, and the rest of the quantum bits flip accordingly:
\begin{equation}
\label{eq:psi1}
|\psi_1\rangle_l=\psi_{\vec d^{(l)}}|1\rangle^{\otimes n}|0\rangle_R+\!\!\!\!\sum_{|\overline{j\oplus \vec{d}^{(l)}}\rangle\neq|1\rangle^{\otimes n}}\!\!\!\!\! \psi_j  | \overline{j \oplus \vec{d}^{(l)}}\rangle|0\rangle_R.
\end{equation}
The overline in the last term indicates that the bit flip occurs if the control bit is $0$. After step 1, the controlled qubit rotation, $C^{n}R_y(\theta^{(l)})$, denoted as $\theta^{(l)}$ in the figure, is applied to the register qubit. The quantum state at step 2 becomes
\begin{equation}
\label{eq:psi2}
|\psi_2\rangle_l=\psi_{\vec d^{(l)}}|1\rangle^{\otimes n}|\theta^{(l)}\rangle_R+\!\!\!\!\sum_{|\overline{j\oplus \vec{d}^{(l)}}\rangle\neq|1\rangle^{\otimes n}}\!\!\!\!\! \psi_j  | \overline{j \oplus \vec{d}^{(l)}}\rangle|0\rangle_R,
\end{equation}
where $|\theta\rangle = \cos\theta|0\rangle+\sin\theta|1\rangle$. The $\bar c X$ gates conditioned on $\vec d^{(l)}$ are applied again to revert the bus state:
\begin{equation}
\label{eq:psi3}
|\psi_3\rangle_l = \psi_{\vec d^{(l)}}|\vec d^{(l)} \rangle|\theta^{(l)}\rangle_R+\!\!\sum_{j\neq\vec d^{(l)}}\!\psi_j |j\rangle|0\rangle_R.
\end{equation}
The second round registers the next data $\vec d^{(l+1)}$ and $\theta^{(l+1)}$:
\begin{equation}
\label{eq:psi4}
|\psi_4\rangle_{l,l+1} = \psi_{\vec d^{(l)}}|\vec d^{(l)} \rangle|\theta^{(l)}\rangle_R+\psi_{\vec d^{(l+1)}} |\vec d^{(l+1)} \rangle|\theta^{(l+1)}\rangle_R+\sum_{j\neq\vec d^{(l)},\vec d^{(l+1)}} \psi_j|j\rangle|0\rangle_R.
\end{equation}
This process can be repeated as many times as the number of data entries. In this way, $M$ data entries can be registered with non-uniform weights to generate a state,
\begin{equation}
\label{eq:psiqram}
\sum_{l=0}^{M-1}\psi_{\vec d^{(l)}}|\vec d^{(l)}\rangle\big{[}\cos\theta^{(l)}|0\rangle_R + \sin\theta^{(l)} |1\rangle_R\big{]}+\!\!\!\sum_{j \notin \{\vec d^{(l)}\} }\!\psi_j  |j\rangle|0\rangle_R.
\end{equation}
Finally, the queried QDB derived from Eq.~(\ref{eq:psiqdb}) can be obtained by selecting an appropriate angle $\theta^{(l)}$ to match the desired probability amplitude $b_l$, and post-selecting the measurement outcome $|1\rangle_R$. The probability to measure $|1\rangle_R$ is
\begin{equation}
    P(1)=\sum_{l=0}^{M-1}|\psi_{\vec{d}^{(l)}}\sin\theta^{(l)}|^2.
\end{equation}
The post-selection increases the total runtime by a factor of $\sim 1/P(1)$, which is data dependent. In some instances, such as in the distance-based quantum classifier~\cite{qClassificationEPL}, the post-selection success probability can be improved by pre-processing the classical data so that $\theta^{(l)}$ is close to $k\pi/2$ for all $l$ where $k$ is an odd-integer.

For the quantum state containing data as equal superposition, the controlled rotations can be replaced with the controlled-NOT gate. Then, the classical data is encoded only in the digital form. If the bus qubit is not in the basis state that corresponds to a data entry $\vec d^{(j)}$, then the $j$th data entry cannot be written in the queried QDB. Moreover, when the same bit string appears more than once, the register qubit accumulates the rotation

Updating desired data entries of the existing quantum database using our scheme is straightforward. The update can be done by inserting the QDB state as the bus qubit and addressing only the target basis states that are to be updated with the selective flip-flop process.

It is important to note that the post-selection process is not always necessary. For example, the post-selection is not needed when all or some of the bus qubit states are addressed to write or modify the binary data to generate the transformation,
\begin{equation}
\text{QRAM}(D)\sum_{j} \psi_j |j\rangle_B |D^0_j\rangle_R = \sum_{j}\psi_j | j\rangle_B |D_j\rangle_R,
\end{equation}
where $D_j^0,\; D_j\in\lbrace 0,1\rbrace$. With $r$ register qubits, this process can be easily generalized to encode $r$-bit data.

In most of the big data applications, real-valued data is encoded in the probability amplitude as discussed thus far. However, if desired, our method can also encode complex probability amplitudes by using a controlled rotation around an arbitrary axis.

In addition to $O(Mn)$ flip-register-flop steps, the total FF-QRAM cost must include the resource overhead for the \textit{register} operations. In fact, the number of elementary gates needed for this step can dominate the runtime of the entire QRAM process. Thus efficient realization of $C^{n}R_p(\theta)$ is critical for the practicality of our scheme. Though the optimal circuit depth reduction can be carried out based on the naturally available set of gates in a specific experimental setup, and is beyond the primary scope of this paper, we briefly mention some examples on how to implement $C^{n}R_p(\theta)$ here. If energy splittings between all pairs of the computational basis states are distinct, then in principle, a resonant pulse at the frequency corresponding to the energy difference between $|1\rangle^{\otimes n}|0\rangle$ and $|1\rangle^{\otimes n+1}$ can realize the desired $C^{n}R_p(\theta)$. But this condition becomes exponentially challenging to satisfy in practice as the number of qubits increases. On the other hand, we can decompose the controlled rotation as $C^{n}R_y(\theta)= $C$^n$NOT~$R_y(\theta/2)$~C$^n$NOT~$R^{\dagger}_y(\theta/2)$. The C$^n$NOT gate can be further decomposed into $2n-3$ Toffoli gates with $n-2$ ancilla qubits prepared in $|0\rangle$ (see Methods). A Toffoli gate can be realized by applying a frequency-selective on-resonance pulse as described above if a set of three qubits is fully addressable while decoupled from the rest of the qubits in the system. Alternatively, a Toffoli gate can be decomposed into five two-qubit gates without requiring ancilla qubits~\cite{PhysRevA.52.3457}. Other methods for implementing C$^{n}$NOT using $O(n)$ number of elementary gates and ancillary space are discussed in refs.~\cite{PhysRevA.52.3457,1255714,PhysRevA.75.022313}. The circuit optimization in terms of Clifford and $T$ gates can be performed using the techniques presented in refs.~\cite{Childs9456,PhysRevX.8.041015}.

\begin{figure}[t]
\centering
\includegraphics[width=0.5\linewidth]{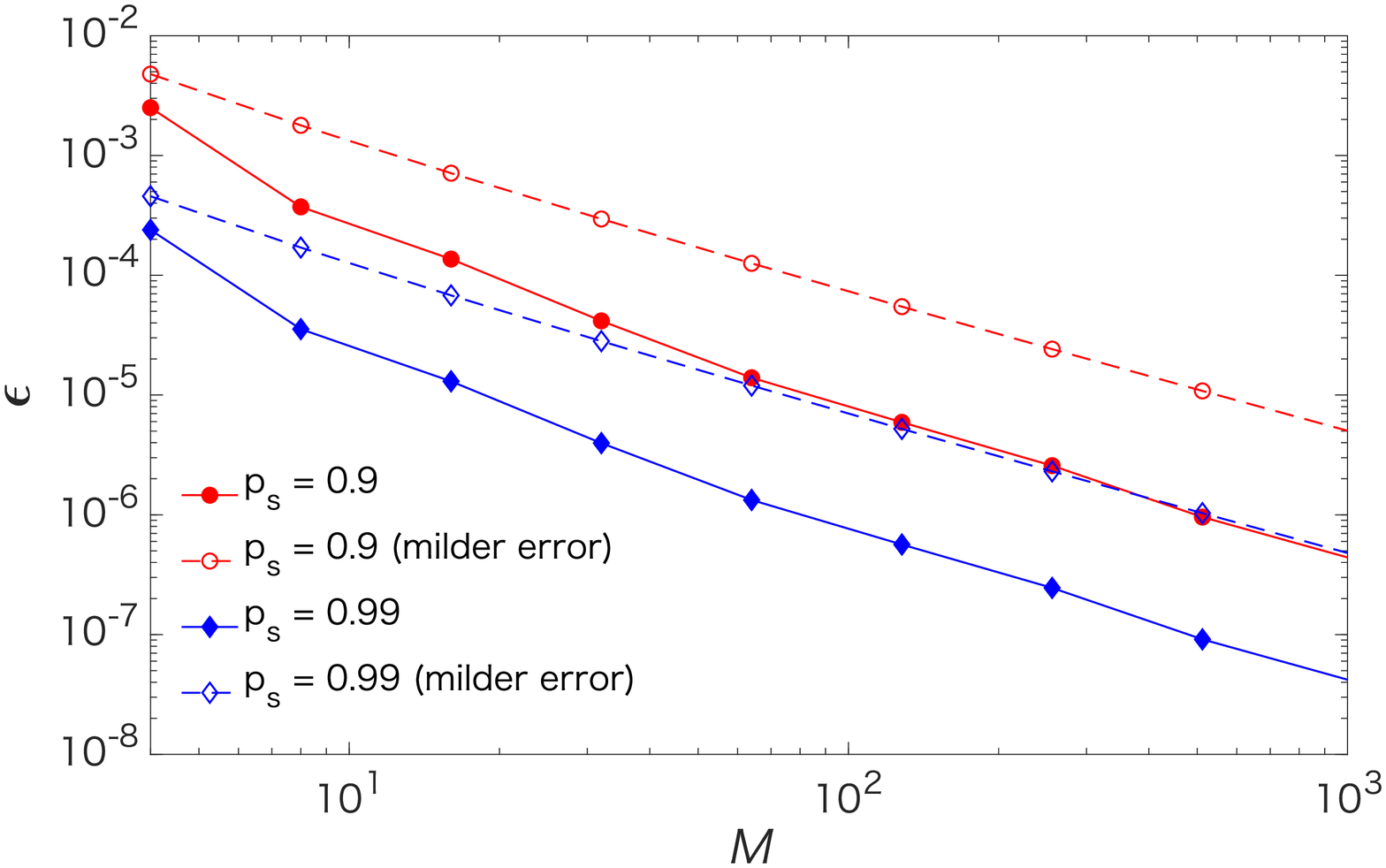}
\caption{Error rate at each time step required for writing $M$ classical data and $n=\log_2(M)$ with the success probability $p_s$. Solid lines with filled symbols are obtained when $C^nR_p(\theta)$ can fail at any of its decomposed locations independently, while dashed lines with open symbols are obtained when the multi-qubit controlled rotation only causes independent errors on $n+1$ qubit, as described in the main text.}
\label{fig:qram_error}
\end{figure}

We investigate the robustness of the FF-QRAM shown in Fig.~\ref{fig:qRAM_fig1} under imperfections using a simple but relevant error model. We assume a typical depolarizing error, in which the state at each time step becomes the maximally mixed classical state with probability $\epsilon$, and remains unchanged with probability $1-\epsilon$. Here, we use the Toffoli gate as an example to count the number of time steps, while further gate decomposition and optimization can be required depending on the experimental setup as mentioned above. When implementing the C$^n$NOT, $2n-1$ qubits undergo $2\lceil\log_2 (n)\rceil-1$ time steps. Therefore, the success probability after writing $M$ classical bit strings of length $n$ with arbitrary probability amplitudes is $(1-\epsilon)^{O(Mn\log_2 (n))}$. As an illustrative example, solid lines in Fig.~\ref{fig:qram_error} shows the individual error rate at each time step necessary for writing $M$ classical bit strings with arbitrary probability amplitudes, assuming $n=\log_2(M)$ without loss of generality, with the success probability $p_s$ of the total QRAM process. A milder assumption that the imperfect $C^nR_p(\theta)$ operation causes independent errors on $n+1$ qubits yields a better success probability, $(1-\epsilon)^{O(Mn)}$. This case is plotted as dashed lines with open symbols in the figure.
The Methods section elaborates on how the number of noisy operations is counted.

Since our scheme is based on the quantum circuit model, fault-tolerant quantum error correction techniques~\cite{PhysRevLett.113.080501,PhysRevLett.111.090505,PhysRevLett.112.010505} can be employed to enhance the accuracy. In contrast, if quantum error correction is applied to BB-QRAM, all routing components are activated at a physical level and make the scheme equivalent to the conventional fanout architecture~\cite{1367-2630-17-12-123010}. In addition, depending on the physical setup, the quantum circuit can be further optimized using various gate decomposition techniques ~\cite{PhysRevA.52.3457,Childs9456,PhysRevLett.93.130502,PhysRevA.93.032318}.

\subsection*{Application to quantum support vector machine}

\begin{figure}[t]
\centering
\includegraphics[width=0.7\linewidth]{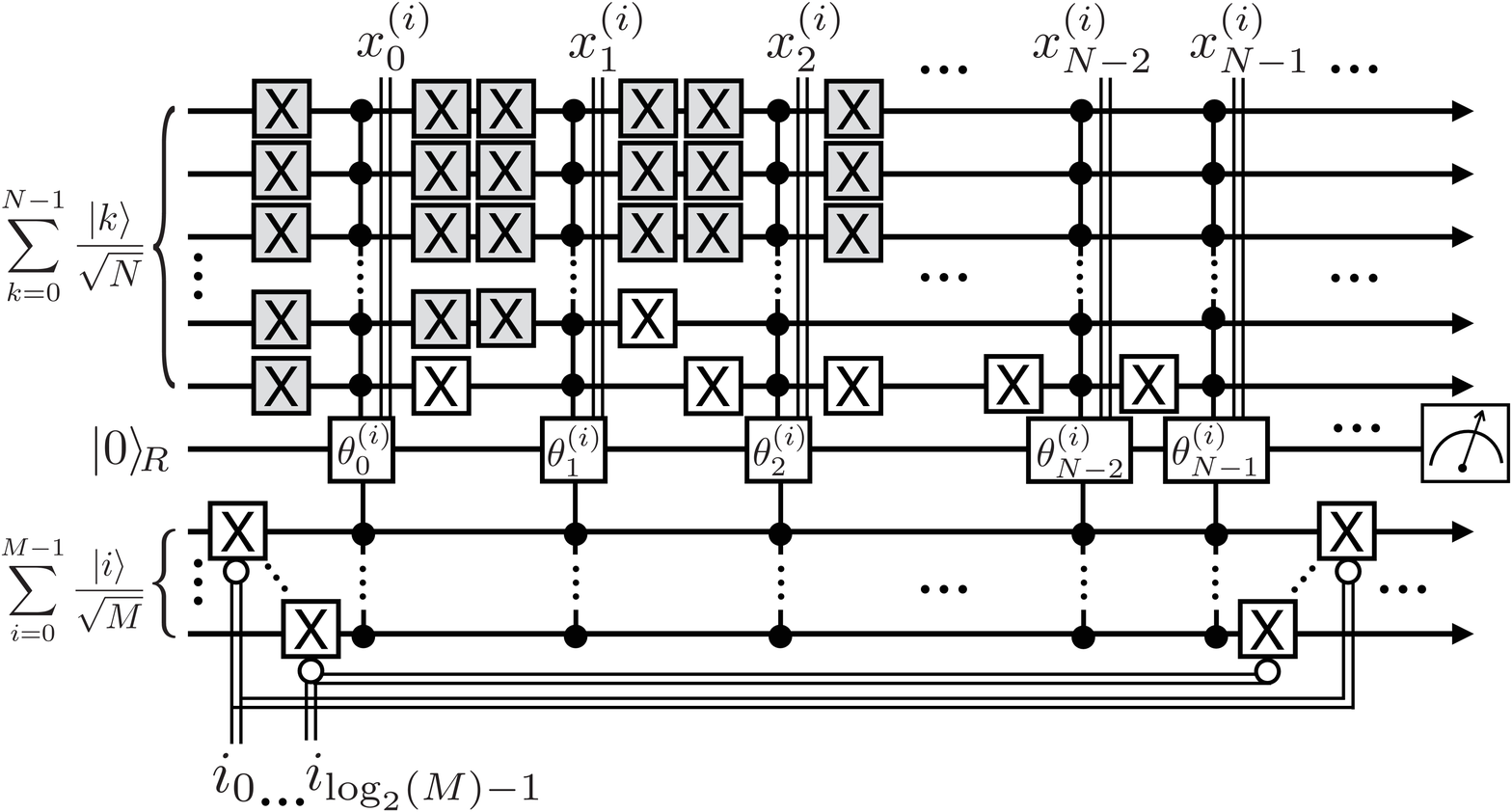}
\caption{Quantum circuit for preparing the QDB for a quantum support vector machine. The gates are shown for writing an $i$th training data only. The gates shaded in gray are added solely for illustrating the flip-flop process, and are not implemented in practice.}
\label{fig:svm}
\end{figure}

As an example, we demonstrate the FF-QRAM process for preparing a training data state in the quantum support vector machine. The classified training examples, $\vec x^{(i)}\in\mathbb{R}^N$, need to be encoded in the quantum format as
\begin{equation}
\label{eq:svm1}
|\chi\rangle=\frac{1}{\sqrt{\sum_{i=0}^{M-1} \lvert\vec x^{(i)}\rvert^2}}\sum_{i=0}^{M-1}\sum_{k=0}^{N-1} x^{(i)}_k|k\rangle |i\rangle,
\end{equation}
where $M$ denotes the number of training data samples~\cite{PhysRevLett.113.130503}. The quantum circuit for preparing this QDB is depicted in Fig.~\ref{fig:svm} for a particular training data with an index $i$, where without loss of generality, $M$ and $N$ are assumed to be powers of two. An equal superposition state for $\log_2(M)$ and $\log_2(N)$ qubits is used as the bus qubit for providing the computational basis states $k$ and $i$, respectively. The $j$th element of $\vec x^{(i)}$ determines the angle $\theta_j^{(i)}$ for the $y$-axis rotation of the register qubit. Note that the gates shaded in gray are included only to show the full flip-flop process for addressing a specific computational basis state, but they can be omitted in the implementation.
The state shown in Eq.~(\ref{eq:svm1}) is obtained after post-selecting the measurement outcome $|1\rangle_{R}$ on the register qubit. This scheme constructs the QDB shown in Eq.~(\ref{eq:svm1}) using $O(\log_2 (MN))$ hardware resources and $O(MN)$ flip-register-flop operations.

\subsection*{Quantum forking}

\begin{figure}[t]
\centering
\includegraphics[width=0.8\linewidth]{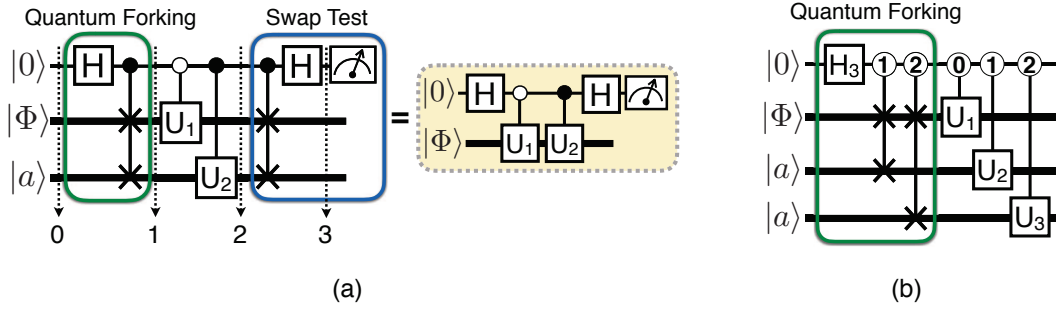}
\caption{(a) Quantum forking to reduce the number of QRAM calls for evaluating the functions of inner products. The swap operation is represented by two $\times$ symbols connected with a vertical line. (b) Quantum forking circuit for evolving the target state $|\Phi\rangle$ under three different unitary operations in each branch (subspace). Thicker horizontal lines indicate multi-qubit channels, $|a\rangle$ is an arbitrary ancilla state, and $H$ $(H_3)$ represents the Hadamard gate on a qubit (qutrit).}
\label{fig:qRAM_fig4}
\end{figure}

Here, we introduce a concept of quantum forking (QF) with which a qubit can undergo independent processes in superposition. This can be utilized as a means to reduce the number of QRAM queries in certain applications. Let us consider a quantum state $|\Psi_0\rangle=|0\rangle|\Phi\rangle|a\rangle$ with an $n$-qubit QDB state $|\Phi\rangle$ generated by a QRAM process and an arbitrary $n$-qubit state $|a\rangle$, where $|\Psi_s\rangle$ denotes the state at step $s$ in Fig.~\ref{fig:qRAM_fig4} (a). An $n$-qubit swap gate between $|\Phi\rangle$ and $|a\rangle$ controlled by a qubit in $(|0\rangle+|1\rangle)/\sqrt{2}$ forms an entangled state,
\begin{equation}
|\Psi_1\rangle=\frac{1}{\sqrt{2}}\left(|0\rangle|\Phi\rangle|a\rangle+|1\rangle|a\rangle|\Phi\rangle\right).
\end{equation}
In other words, the QDB is encoded in the first $n$-qubit data block if the control qubit is $0$, and in the second $n$-qubit data block if the control qubit is $1$. Then by applying two unitary evolutions activated by different computational basis states of the control qubit to each $n$-qubit block, $|\Phi\rangle$ forks into two different states in superposition:
\begin{equation}
\label{eq:QF}
|\Psi_2\rangle=\frac{1}{\sqrt{2}}\left(|0\rangle U_1|\Phi\rangle |a\rangle+|1\rangle |a\rangle U_2|\Phi\rangle\right).
\end{equation}

Evidently, it is not possible to create correlations between $|\Phi_1\rangle=U_1|\Phi\rangle$ and $|\Phi_2\rangle=U_2|\Phi\rangle$ via linear operations. Nonetheless, QF can speedup certain tasks, such as ensemble averaging~\cite{QFforFastSampling} and the inner product calculation. Here we focus on the inner product evaluation problem as an example. The inner product between $|\Phi_1\rangle$ and $|\Phi_2\rangle$ can be evaluated by preparing these two states individually by making queries to the QRAM and performing the swap test~\cite{PhysRevLett.87.167902}. Alternatively, starting from the state shown in Eq.~(\ref{eq:QF}), another controlled swap followed by a Hadamard operation on the control qubit yields the state
\begin{equation}
|\Psi_3\rangle=\frac{1}{2}\big{[}|0\rangle\left(|\Phi_1\rangle+|\Phi_2\rangle\right)+|1\rangle\left(|\Phi_1\rangle-|\Phi_2\rangle\right)\big{]}|a\rangle.
\end{equation}
Finally, the probability of measuring the control qubit in $|0\rangle$ is given by $P(0)=[1+\text{Re}(\langle \Phi_1|\Phi_2\rangle)]/2$. This procedure only reveals the real part of the inner product. The imaginary part can be evaluated by adding a phase gate to the control qubit in front of the final Hadamard gate. Since the ancilla qubit is in an arbitrary state, QRAM is used only for preparing $|\Phi\rangle$ once. The ancilla qubit can even be in the maximally mixed state, and we assume that the cost of preparing such a state is negligible. This method consumes $O(n)$ additional gates, but reduces the number of QRAM queries by a factor of $\sim 1/2$.
Note that the conventional swap test can only estimate the magnitude of the inner product. Thus the QF based approach not only reduces the number of QRAM queries, but also allows for the determination of the sign of the inner product. This is a consequence of an important property of the QF circuit; since different unitary operators can be applied to each branch (subspace), the global phase that a unitary operator introduces become distinguishable. Clearly, the quantum circuit shown on the left side of Fig.~\ref{fig:qRAM_fig4} (a) can be rewritten more compactly without the controlled swap gates and the ancilla qubit by applying both controlled unitary operators directly to the data qubit as shown on the right side. The quantum circuit on the left illustrates the general quantum forking framework that can also be adapted for other applications by replacing the swap test with other measurement schemes~\cite{QFforFastSampling}.
 
Generalizing above idea, a quantity such as $\sum_{1\le i,j\le d}\text{Re}(\langle \Phi_i|\Phi_j\rangle)$ can be evaluated by repeating the modified swap test for which only one QRAM state preparation is needed. The modified swap test based on QF requires a control qudit of dimension $d$ (or $\log_2 (d)$ qubits), and $O(nd)$ gates.

The quantum forking for implementing three different unitary processes in superposition is depicted in Fig.~\ref{fig:qRAM_fig4} (b). A qutrit is used as the control, and $H_3$ represents the qutrit Hadamard operation for preparing the equal superposition of the three computational basis states. This circuit produces an entangled state $(|0\rangle|\Phi_1\rangle|a\rangle|a\rangle+|1\rangle|a\rangle|\Phi_2\rangle|a\rangle+|2\rangle|a\rangle|a\rangle|\Phi_3\rangle)/\sqrt{3}$.

\section*{Discussion}

Encoding large classical data into a quantum database must be done efficiently in a way that the potential advantages of the quantum algorithms for big data applications do not vanish. We proposed the flip-flop QRAM, a systematic architecture, for preparing a quantum database using the quantum circuit model. The circuit-based construction is imperative since it provides flexibility and compatibility with existing quantum computing techniques. Our process can register $n$-bit classical data with arbitrary probability amplitudes stored in $M$ memory cells into quantum format using $O(n)$ qubits and $O(Mn)$ flip-register-flop steps. The versatility of the architecture allows to create a complex data structure via encoding any classical information, either discrete or continuous, as quantum bits or as probability amplitudes of a quantum state. An example of the amplitude encoding is the  application to a quantum state generation for a quantum support vector machine algorithm in which the training data is represented with the probability amplitudes as shown in Fig.~\ref{fig:svm}. Qubit encoding can be achieved by beginning with the quantum bus state as $|\psi\rangle_B = |+\rangle^{\otimes n}$, and inserting the weights (e.g., the normalized occurrence) of the data by adjusting the multi-qubit controlled rotation $C^nR_p(\theta)$. For the uniform weight which is, for example, encountered in the parity learning algorithm~\cite{LPNTheory,LPNexp}, the multi-qubit controlled gate is simply $C^n\text{NOT}$. For the amplitude encoding, the final QDB state is obtained by post-selecting on the register qubit being $|1\rangle_R$. Hence the amplitude encoding introduces additional resource overhead for repeating the entire algorithm. However, for some tasks, the classical data can be pre-processed to increase the success probability of the post-selection. It is an interesting open problem whether the post-selection can be avoided in certain amplitude encoding schemes by utilizing the fact that the probability amplitudes are determined by cosines instead of sines if the register qubit is $|0\rangle_R$. Also, the post-selection can be avoided for the qubit encoding if the bus qubit state only contains the basis states that are to be queried and all classical data are registered with an equal weight. Note that BB-QRAM also employs the post-selection for preparing an arbitrary QDB state with the amplitude encoding. With some limitations, the amplitude encoding can be done without relying on the post-selection. For example, ref.~\cite{PhysRevX.8.041015} introduces a procedure inspired by classical alias sampling to assign the probability amplitude $\sqrt{\tilde{\rho}}$ using $2\mu+2n+O(1)$ ancilla qubits, where $\tilde{\rho}$ is $\mu$-bit binary approximation to the desired non-negative real value, $\rho$. In ref.~\cite{PhysRevA.97.052329}, adiabatic-diabatic state preparation is used to generate superposition states with squared amplitudes.

We point out potential solutions to several issues for meaningful applications of QRAM. First, when the FF-QRAM process leaves the last term in Eq.~(\ref{eq:psiqram}) that corresponds to the states without data entries, the rate of producing the desired QDB can be reduced. This issue can be partially circumvented by running $L$ identical QRAM processes in parallel. Then the success probability of the post-selection improves by a factor of $L$ while also increasing the number of qubits and the gates by the same factor. Note that the time complexity remains the same. Second, the QDB is not reusable once it is consumed by a quantum algorithm since the measurement collapses the state. Motivated by the above, we introduced the concept of quantum forking that allows to reduce the number of QRAM queries in some instances, in particular, when evaluating the inner product. Finding other instances for which the quantum forking can reduce the number of QRAM calls, even by a constant amount, remains an interesting open problem.

\section*{Methods}
\subsection*{Error analysis}
\begin{figure}[t]
\centering
\includegraphics[width=0.5\linewidth]{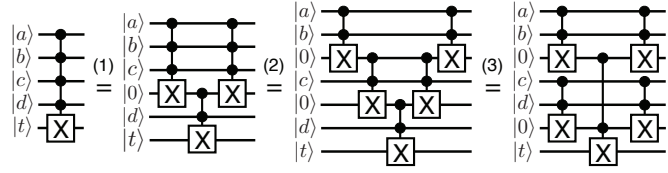}
\caption{Quantum circuit for implementing a C$^4$NOT gate using $5$ Toffoli gates and $2$ ancilla qubits prepared in $|0\rangle$ in $3$ steps. The control bits are $a,\; b,\; c$ and $d$, and the target qubit is $|t\rangle$. The gates after the Toffoli gate for conditionally flipping the target qubit uncompute the ancila qubits in order to unentangle the ancillae from the system.}
\label{fig:CnNOT}
\end{figure}

The C$^n$NOT gate can be decomposed into a C$^{n-1}$NOT and a C$^{2}$NOT (Toffoli) using an ancilla qubit prepared in $|0\rangle$ as shown in step (1) of Fig.~\ref{fig:CnNOT} for $n=4$ as an example. Then by recursion, C$^n$NOT can be realized using $2n-3$ Toffoli gates and $n-2$ ancilla qubits prepared in $|0\rangle$ (step (2) in Fig.~\ref{fig:CnNOT}). Note that $n-2$ Toffoli gates are added after the Toffoli gate for conditionally flipping the target qubit in order to unentangle the ancilla qubits from the system. The quantum circuit can be rearranged to further reduce the depth as shown in the last step in Fig.~\ref{fig:CnNOT}.

We assume that the $\bar c X$ operations can occur simultaneously on all target qubits, but even when the control bit is 1, the target qubit may undergo an erroneous identity operation. To reduce the circuit depth, the $\bar c X$ operations after $C^{n}R_y(\theta^{(l)})$ and before $C^{n}R_y(\theta^{(l+1)})$ can be merged. The combined operation flips the $j$th qubit only if $d_j^{(l)}\oplus d_j^{(l+1)}=1$, and otherwise does nothing, for $j=0,\ldots, n-1$. Thus, the total number of single qubit gates used for writing $M$ classical data of length $n$ is $n(M+1)$. Each $C^{n}R_y(\theta)$ uses two single qubit gates and two C$^n$NOT gates.
In the C$^n$NOT implementation described above (Fig.~\ref{fig:CnNOT}), $2n-1$ qubits ($n$ control qubits $+\; 1$ target qubit $+\; n-2$ ancillae) undergo $2\lceil\log_2 (n)\rceil - 1$ time steps.

Therefore, the total number of time steps $\tau$  that are subject to noise can be counted as
\begin{equation}
\tau=2M\big{[} (2n-1)(2\lceil\log_2 (n)\rceil-1)+1\big{]}+n(M+1).
\end{equation}
If $C^nR_y(\theta)$ can be implemented with only $n+1$ independent errors, then $\tau$ can be further reduced to $(n+1)M+n(M+1)$.

\subsection*{Data availability}
The datasets generated during and/or analysed during the current study are available from the corresponding author on reasonable request. 

\section*{Acknowledgements}
This research is supported by the National Research Foundation of Korea (Grant No. 2016R1A5A1008184), by the MSIT, Korea, under the ITRC support program (IITP-2018-2015-0-00385 and IITP-2018-2018-0-01402), by the KI Science Technology Leading Primary Research Program of KAIST, and by the South African Research Chair Initiative of the Department of Science and Technology and the National Research Foundation. We thank Maria Schuld and Carsten Blank for stimulating discussions.

\section*{Author contributions statement}

J.K.R. conceived the project. All authors contributed towards the design and analysis of the model. D.K.P. and J.K.R. wrote the main manuscript text. All authors reviewed the manuscript.

\section*{Additional information}

\textbf{Competing interests} The authors declare no competing interests.

%\bibliography{ref}
\end{document}